\documentclass[epjCONF, onecolumn]{svjour}
\usepackage{graphics}
\usepackage[varg]{txfonts} 
\usepackage[latin1]{inputenc}
\newcommand\kms{{\rm\,km\,s^{-1}}}

\newcommand\zmax{z_{\rm max}}
\newcommand\ulsr{U_{\rm LSR}}
\newcommand\vlsr{V_{\rm LSR}}
\newcommand\wlsr{W_{\rm LSR}}

\session-title{Assembling the puzzle of the Milky Way}
\begin{document}
\title{Chemical constraints on the formation of the~Galactic~thick~disk}
\author{Thomas~Bensby \and Sofia~Feltzing}
\institute{Lund Observatory, Department of Astronomy and Theoretical Physics, Lund, Sweden}
%
\abstract{
We highlight some results from our detailed abundance analysis study 
of 703 kinematically selected 
F and G dwarf stars in the solar neighbourhood. The analysis is based on spectra of
high-resolution ($R=45\,000$ to 110\,000) and high signal-to-noise 
($S/N\approx 150$ to 300). The main findings include: (1) at a given
metallicity, the thick disk abundance trends are more $\alpha$-enhanced than those
of the thin disk; (2) the metal-rich limit of the thick disk reaches at least 
solar metallicities; (3) the metal-poor limit of the thin disk is around 
$\rm [Fe/H]\approx-0.8$; (4) the thick disk shows an age-metallicity gradient;
(5) the thin disk does {\it not} show an age-metallicity gradient; (6) the most
metal-rich thick disk stars at $\rm [Fe/H]\approx 0$ are significantly older than the
most metal-poor thin disk stars at $\rm [Fe/H]\approx -0.7$; (7) based on our elemental 
abundances we find that kinematical
criteria produce thin and thick disk stellar samples that are biased in the sense
that stars from the low-velocity tail of the thick disk are classified as thin disk
stars, and stars from the high-velocity tail of the thin disk are classified 
as thick disk stars; (8) age criteria appears to produce thin and thick
disk stellar samples with less contamination. 
} 
\maketitle
%
\section{Introduction}

Thick disks as unique entities in galaxies were discovered in the late 1970s when it was 
found that the vertical light profiles of a few edge-on galaxies could 
not be fitted by single exponentials \cite{burstein1979}. Similarly, the 
Galactic thick disk as a unique stellar population 
was detected in the early 1980:s when star count data towards the Galactic 
South Pole could not be fitted with just one power law but needed two: one with a 
scale-height of 300\,pc and one with a scale-height of 1350\,pc \cite{gilmore1983}. 
The former was associated with the already known thin disk and the latter with 
the newly discovered thick disk. As there is no 
a priori reason that says that vertical star counts in galaxy disks 
must fit single power laws this finding was necessary but 
not sufficient to define the thick disk as a unique entity. 
Since then it has been shown that the Galactic thin and thick disks are also
distinct disk populations in terms of kinematics as well as chemistry:
the thick disk is a more slowly rotating 
stellar system than the thin disk, lagging the local standard of rest (LSR) 
by 40 to 50\,$\kms$ \cite{soubiran2003}; it is older than the thin
disk \cite{fuhrmann1998,bensby_amr}; it has a lower average metallicity
than the thin disk \cite{gilmore1995}; and, at a given metallicity,
it is more $\alpha$-enhanced than the thin disk 
\cite{fuhrmann1998,reddy2006,bensby2003,bensby2005,bensby2007letter2}.
These differences point to separate origins and formation histories of 
the two disks. Also, it has recently been found that the Galactic bulge has a bimodal
metallicity distribution \cite{bensby2011,hill2011} and that the metal-poor part of the
bulge is very similar to the thick disk (average metallicity,
elemental abundance trends, and age distribution) \cite{bensby2010,bensby2011,alvesbrito2010}. This might suggest that the
bulge and the thick disk are tightly connected.

From the Geneva-Copenhagen Survey (GCS)
\cite{nordstrom2004,casagrande2011},
which contains kinematics, ages, and metallicities estimated from
Str\"omgren uvby$\beta$ photometry for
$\sim 14\,000$ nearby dwarf stars,
it is evident that stars with orbital rotational velocities 
typical for the thick disk can be found at very
high metallicities, even well above solar (see Fig.~\ref{fig:uvw_feh}). 
The question is if these stars
are true thick disk stars? Indeed, there has been studies that claim that 
the most metal-rich thick disk stars lies around $\rm [Fe/H]\approx -0.3$
and that the metal-richer ones are likely to be thin disk
stars due to their low $\rm [\alpha/Fe]$ ratios 
\cite{mishenina2004,fuhrmann2004,reddy2006}. However, recent studies
show that the thick disk can be traced to at least solar metallicities
\cite{bensby2007letter2,reddy2010}.

To further investigate the chemical and kinematical properties of the 
Galactic disk, and the degree of the separation between the two disks, 
we have undertaken a large spectroscopic survey
of F and G dwarf stars in the Solar neighbourhood.
The stellar sample consists of 703 stars and is the joint effort from 
several observing projects, each with its own specific goal, aiming at probing 
the different stellar populations in the Solar neighbourhood at their 
extremes. Important scientific questions addressed in our projects include:
(i) How metal-rich can the thick disk be? (ii) How metal-poor 
can the thick disk be? (iii) How metal-poor can the thin disk be?
(iv) How metal-rich can the halo be? (v) Where do the old and metal-rich 
stars come from? (vi) Where do kinematical substructures such as the
Hercules stream and the Arcturus moving group come from? First results 
regarding points (iii) and (vi) can be found in \cite{bensby2007letter2} and 
\cite{bensby2007letter}.
Bringing these questions together will help us to constrain the formation history
of the Milky Way, and will ultimately
add pieces to the puzzle of galaxy formation.

\section{Observations and abundance analysis}

\begin{figure}
\centering
\resizebox{0.95\hsize}{!}{\includegraphics{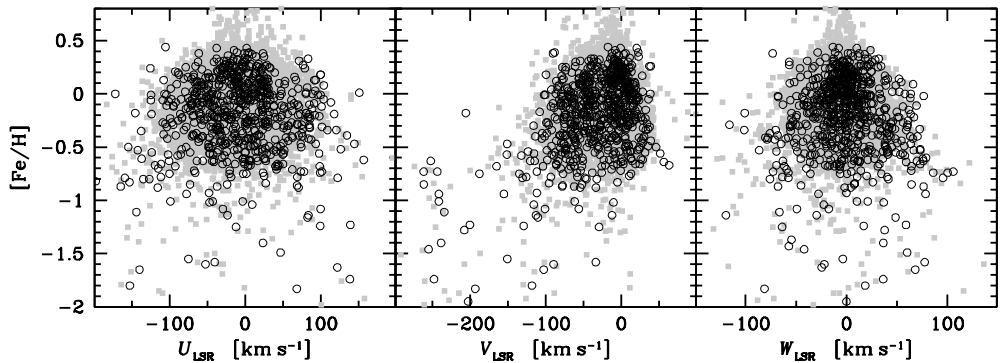}}
\caption{[Fe/H] versus $\ulsr$, $\vlsr$, and $\wlsr$ velocities for our stars (o).
The grey dots in the background represent the $\sim 14000$ stars
in the GCS catalogue using the new metallicities from \cite{casagrande2011}.
\label{fig:uvw_feh}
}
\end{figure}

The stars were selected mainly from the GCS and a few, that were
not available in the GCS, from the
catalogue by Feltzing \& Holmberg \cite{feltzing2001}. 
Figure~\ref{fig:uvw_feh} shows [Fe/H] as a function of the 
$\ulsr$, $\vlsr$, and $\wlsr$ velocities. One aspect of our sample that sets it apart 
from, e.g., the studies by Reddy et al. \cite{reddy2003,reddy2006}, is that we cover 
a wider range of orbital parameters.  
For example our sample includes stars on orbits with low 
eccentricity {\sl and} low [Fe/H] 
(chosen to study the metal-weak thin disc) as well as stars with 
super-solar [Fe/H] {\sl and} highly eccentric orbits and/or highly
negative $\vlsr$ velocities (chosen to study the 
metal-rich thick disc). Neither of these types of stars have been 
systematically included in previous studies. In fact in some studies 
they are lacking all together and the inclusion of them in our study 
enables us to explore a wider range of the parameter space.
Figure~\ref{fig:uvw_feh} shows that we explore the full [Fe/H]-velocity plane
of the GCS with high-resolution spectroscopy.

The methods to determine the stellar parameters and to calculate elemental
abundances follow the methodology described in our previous papers that
contain a subset of 102 stars of the current sample
\cite{bensby2003,bensby2005}. Further descriptions regarding the analysis,
updated error estimation methods, and new age determination,
will be included in an upcoming paper (Bensby, Feltzing \& Oey, in prep.).

\begin{figure*}
\centering
\resizebox{0.95\hsize}{!}{
\includegraphics{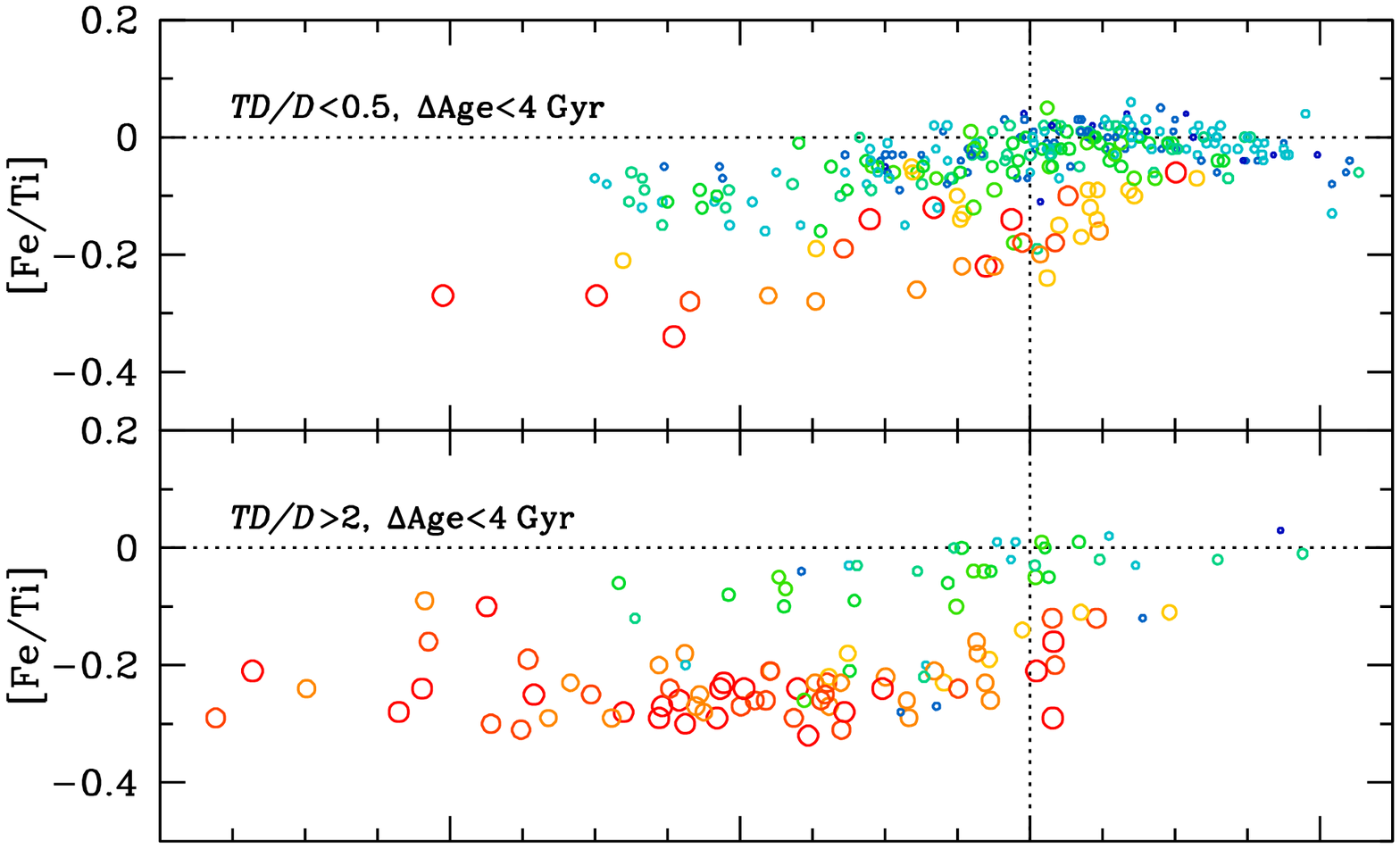}
\includegraphics{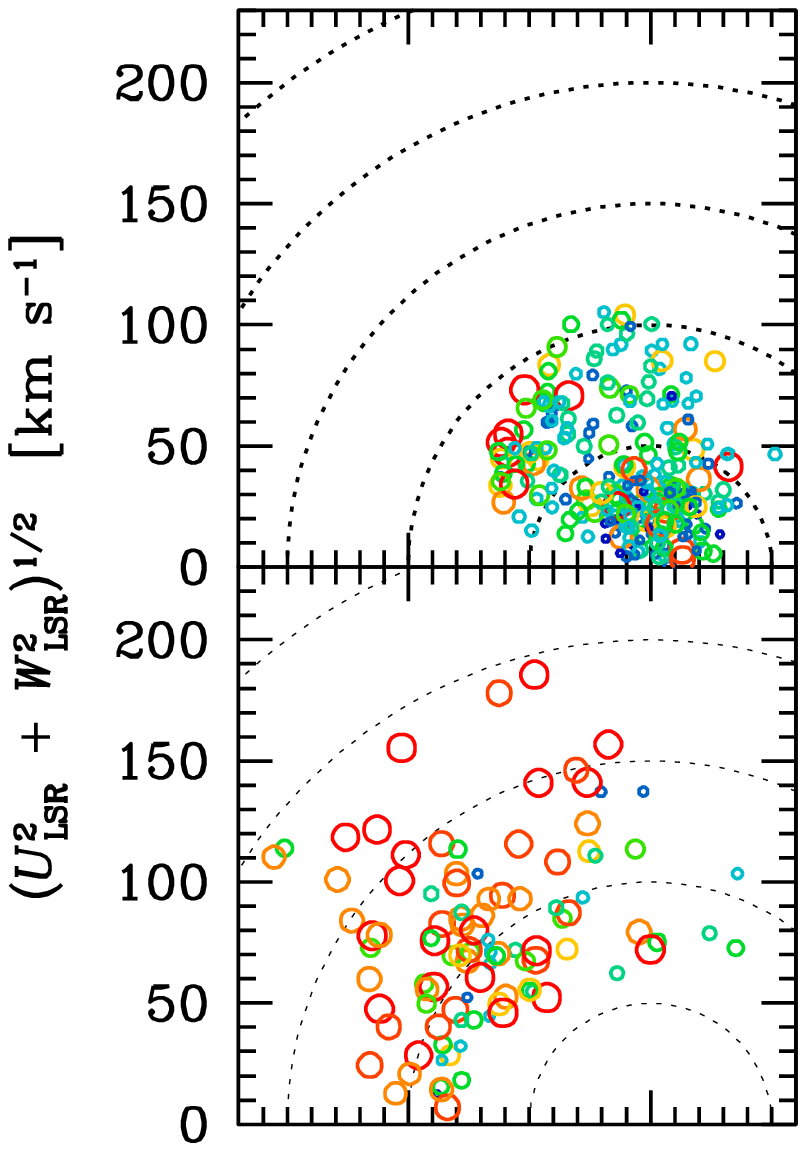}}
\resizebox{0.95\hsize}{!}{
\includegraphics{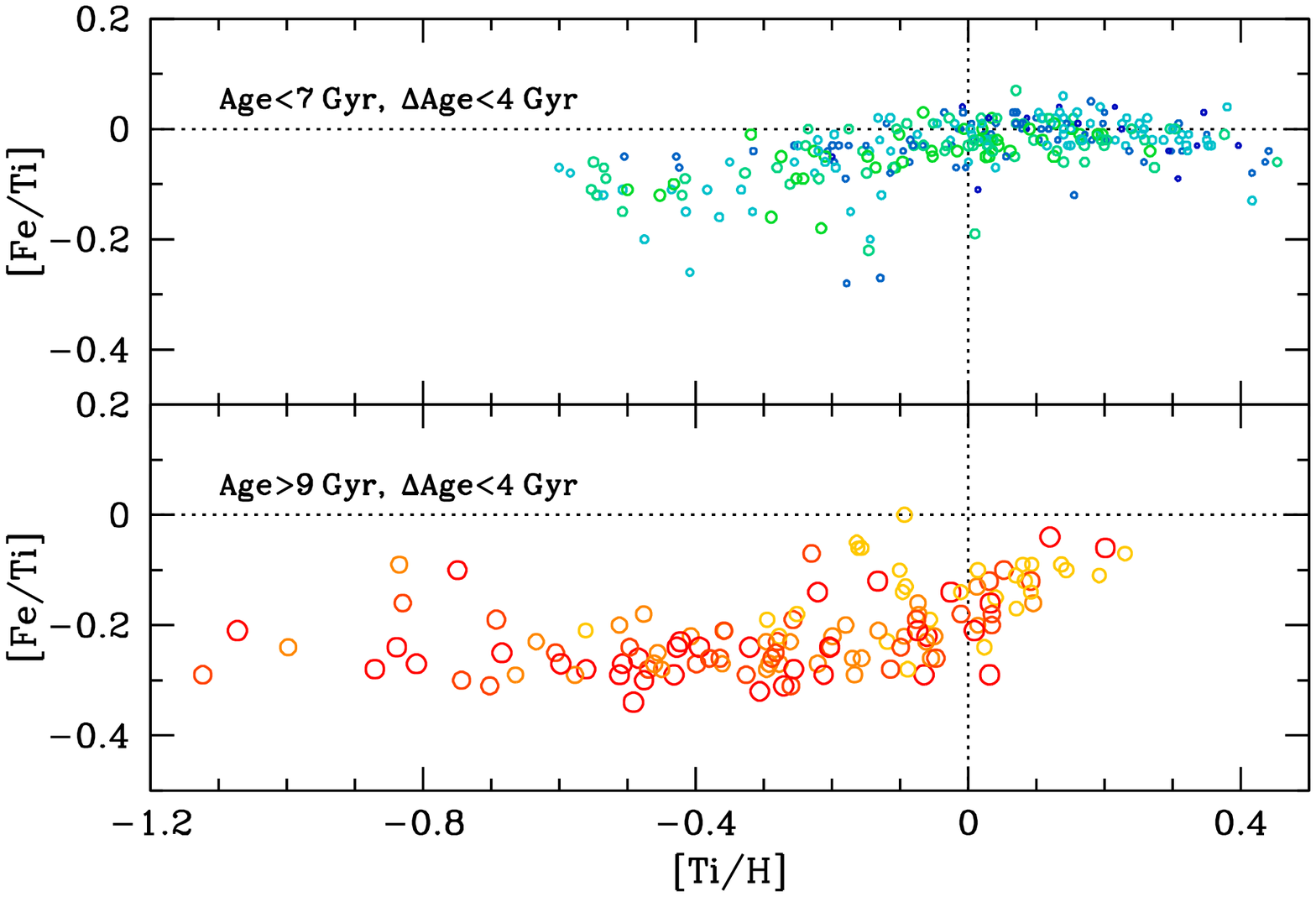}
\includegraphics{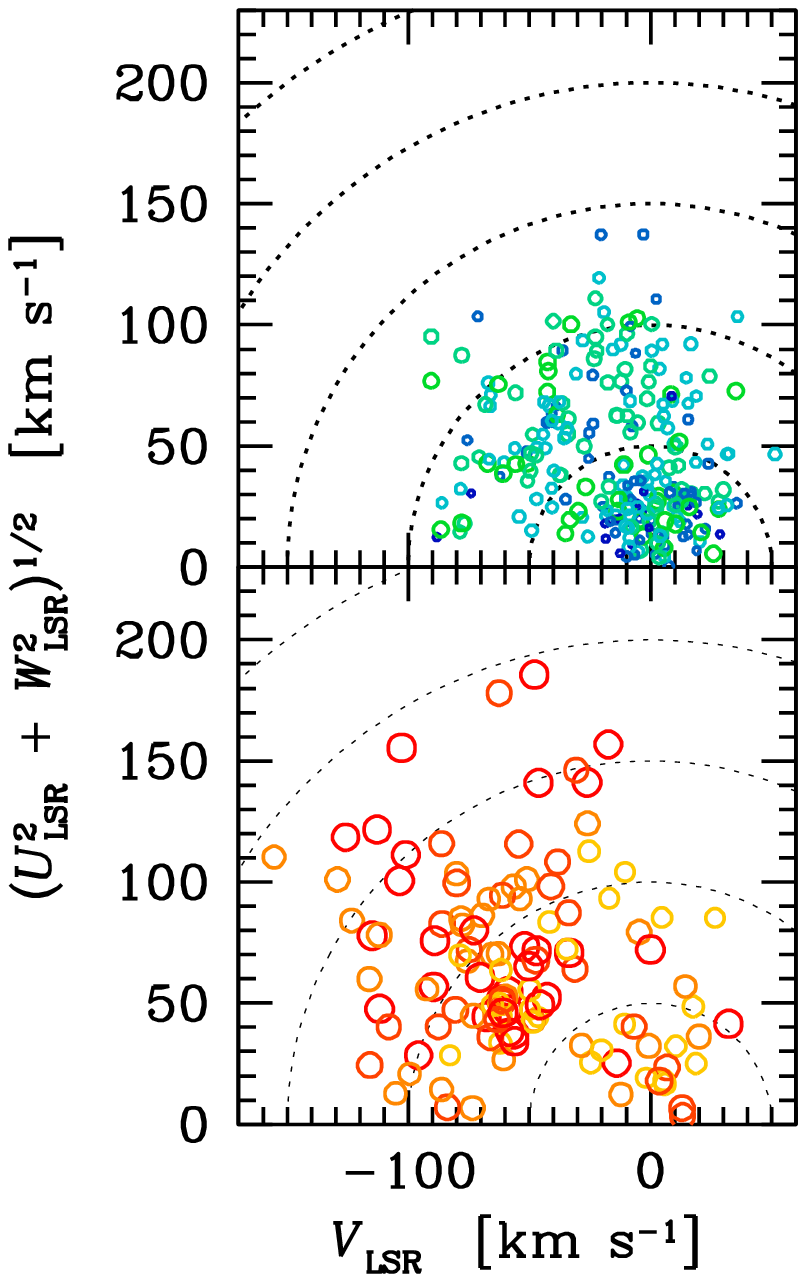}}
\caption{{\sl Upper left:} [Fe/Ti]-[Ti/H] abundance
trends when using the kinematical criteria from 
\cite{bensby2003,bensby2005}; {\sl lower left:} when using
age criteria. {\sl Right:} Toomre diagrams for the samples
on the left-hand side. Only stars with a
difference between upper
and lower age estimates less than 4\,Gyr are shown. 
Stars have been colour- and size coded
depending on their ages (blue and small circles represent young 
stars, red and big circles
represent old stars). 
\label{fig:tife}
}
\end{figure*}

\section{Results and discussion}
\subsection{Selection criteria}

Thick disk stellar candidates have usually been selected based
on kinematical criteria \cite{bensby2003,bensby2005}.
The reason to use kinematical criteria is because you want to investigate
the chemical and age properties of the stellar populations, and if
chemical and/or age criteria are invoked those investigations
would be undermined. Using kinematical criteria,
it has generally been found that stars with thick disk kinematics
are more enhanced in the $\alpha$-elements
than the stars with thin disk kinematics at a give metallicity 
\cite{reddy2003,bensby2003,mishenina2004,bensby2005,reddy2006}, 
and that they are older at all metallicities, even where the metallicities are
over-lapping \cite{fuhrmann1998,fuhrmann2008,bensby2007letter2}.

In the upper left plot in Fig.~\ref{fig:tife} we show the
[Fe/Ti]-[Ti/H] abundance trends for two kinematically selected samples;
one where the probabilities of being a thin disk star is at least twice
that of being a thick disk star $(TD/D<0.5)$, and one where the
probabilities of being a thick disk star is at least twice that of being
a thin disk star $(TD/D>2)$. As can be seen the two 
samples show very different abundance trends. However, it is also 
clear that the main chemical signature for each kinematical sample
appears to be present also in the other sample. We have therefore
also in Fig.~\ref{fig:tife} coded each star by its estimated age. 
It is evident that the weaker signature
in each kinematical sample has the same age structure as the main signature
in the other sample, i.e. the $\alpha$-enhanced stars in the thin disk
sample with $TD/D<0.5$ are older than the ones that are not as 
enhanced, and have the same ages as the $\alpha$-enhanced stars in the
thick disk sample with $TD/D>2$. Vice versa, the stars with low
$\alpha$-enhancements in the thick disk sample are younger than 
the ones with higher $\alpha$-enhancements, and have ages comparable
with the bulk of stars in the thin disk sample. The Toomre diagrams on 
right-hand side of Fig.~\ref{fig:tife} show that
the two samples are kinematically different, with only little
overlap.

The [Fe/Ti]-[Ti/H] abundance trends for two samples, one
old sample with stars that have estimated ages greater than 9\,Gyr,
and one young sample with stars that have estimated age less than
7\,Gyr, are shown in the bottom left plots of Fig.~\ref{fig:tife}.
Once again we see two very different chemical signatures,
similar to the ones where the stars were separated based on their 
kinematics. However, the abundance trends are 
now much cleaner, with essentially no overlap between the two.
Looking at the Toomre diagrams for these two samples (bottom right plots
of Fig.~\ref{fig:tife}) we see that there is a large kinematical
overlap between the two. There are young stars with hot 
kinematics and old stars with cold kinematics. 

It appears that kinematical selection criteria
introduces a ``kinematical confusion" between thin and thick disk 
stars. Due to the different metallicity distributions of the 
thin and thick disks, the problem is especially severe at high metallicities, 
where stars from the 
high-velocity tail of the thin disk are likely to be mis-classified as thick disk
stars, and at low metallicities, where stars from the low-velocity tail of the
thick disk are likely to be mis-classified as thin disk stars.
Instead, stellar ages appear to be able to produce ``cleaner" thin and thick
disk samples. However, as stellar ages require knowledge of stellar parameters
and metallicities, and are also only possible to determine for dwarf and subgiant
stars, kinematics might be the only available discriminator. In those cases,
one should be aware of the possible kinematical confusion between thin and 
thick disk stellar samples.

\subsection{Ages and metallicities}

\begin{figure}
\centering
\resizebox{\hsize}{!}{
\includegraphics{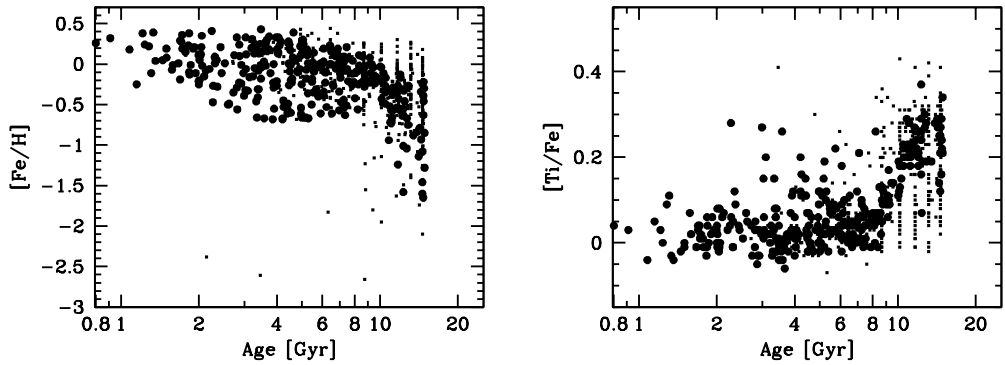}}
\caption{[Fe/H] and [Ti/Fe] versus age. Stars with well-determined
ages
(difference between upper and lower age estimate less than 3\,Gyr)
are marked by larger dots.
\label{fig:agefe}
}
\end{figure}

Figure~\ref{fig:agefe} show the age-metallicity plot for the
stars with estimated ages. We highlight those stars that
have well-determined ages, i.e., age difference between upper and lower 
estimates smaller than 3\,Gyr.
A few noticeable features are: (1) There are no young and metal-poor stars;
(2) There are old and metal-rich stars; (3) The old and metal-rich stars
all have ages with large error bars; (3) For stars with well-determined
ages greater than $\approx 8$\,Gyr, there appears to exist
an age-metallicity relation. 

Also, from Fig~\ref{fig:tife} we see that the most metal-rich stars
with thick disk kinematics are significantly older than the most metal-poor 
stars with thin disk kinematics. Whether or not there is a real hiatus
in the star formation history between the thin and thick disks is unclear,
and requires further investigation with an un-biased sample. However, it is
likely that the Milky Way, after star formation in the thick disk ceased 
(at $\rm [Fe/H]\approx 0$), 
acquired gas of lower metallicity (merger?). Once star formation started
again, it did so in the thin disk at lower metallicities 
($\rm [Fe/H]\approx-0.7$).

\subsection{Variation with galactocentric distance}

\begin{figure*}
\centering
\resizebox{\hsize}{!}{
\includegraphics{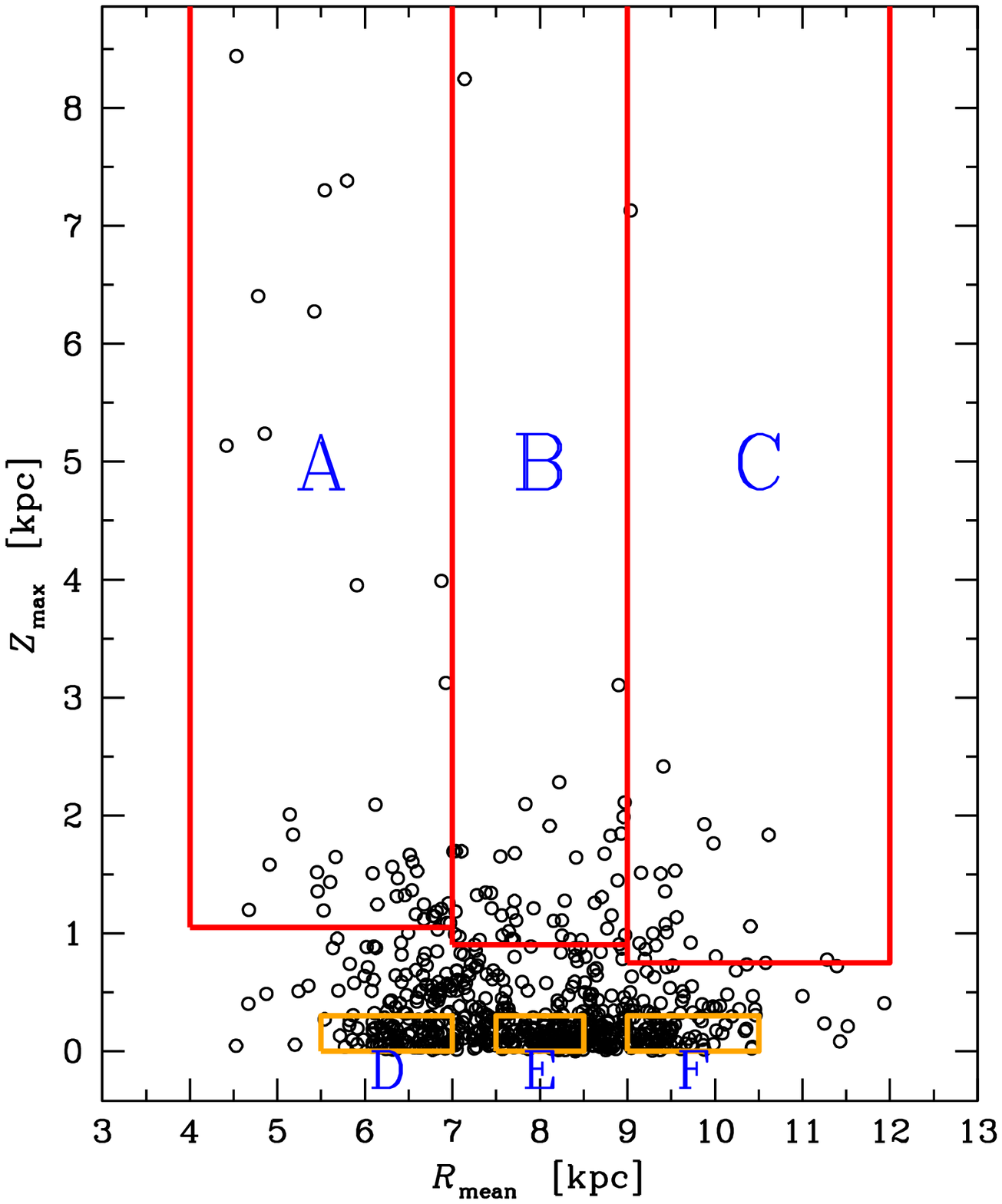}
\includegraphics{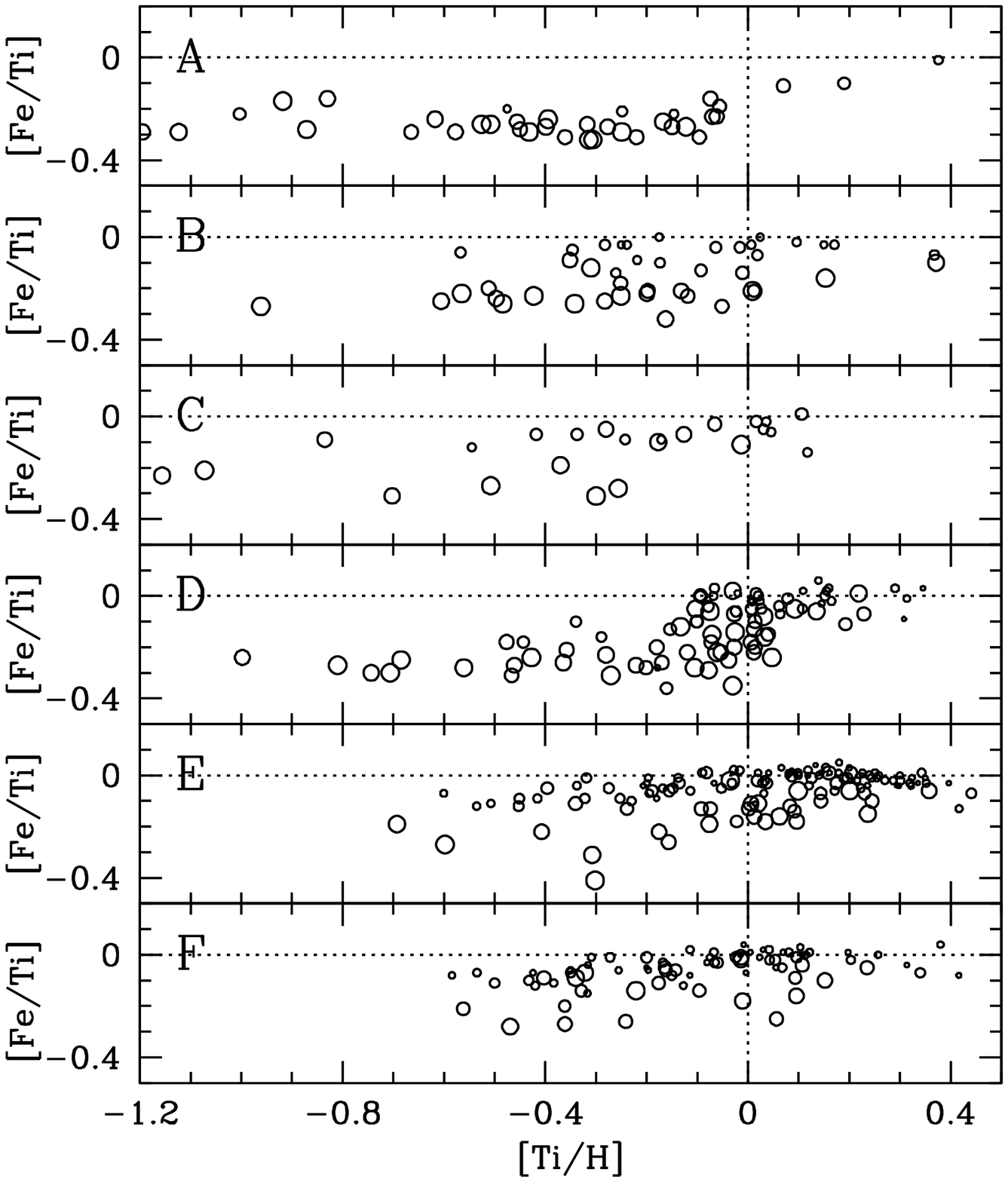}}
\caption{{\sl Left:} $\zmax$ versus $R_{\rm mean}$.  {\sl Right:} [Fe/Ti]-[Ti/H]
for stars with orbital parameters as shown in boxes A-F
in the plot on the left-hand side. The sizes of the circles have been scaled 
with the estimated ages of the stars.
\label{fig:rmeanzmax}
}
\end{figure*}

Figure~\ref{fig:rmeanzmax} shows $\zmax$ as a function of $R_{\rm mean}$ for the
stellar sample. $R_{\rm mean}$ can be regarded as an estimate of the distance from the Galactic
centre where the stars were born \cite{edvardsson1993}. Boxes~A, B, and C contain
stars on orbits that reach far from the Galactic plane, presumably born in
the inner disk, at the solar distance, and in the outer disk, respectively, while
Boxes~D, E, and F contain stars on orbits close
to the plane, presumably born in the inner disk, at the solar distance, and in
the outer disk, respectively. The right-hand side plot of 
Fig.~\ref{fig:rmeanzmax} shows the [Fe/Ti]-[Ti/H] plots for the stars in these
different boxes. The sizes of the markers have been scaled with the ages of the
stars (older stars -- bigger circles). First, we note that stars born at the solar
distance (Boxes B and E) divide into the typical abundance trends we associate
with the thin and thick disks. Box B contains more stars with thick disk
properties (old and $\alpha$-enhanced) while Box~E contains more stars 
with thin disk properties (young and no $\alpha$-enhancement). This is expected.
If the relative properties of the inner and outer disk were similar to what is 
seen in the solar neighbourhood (Box E), the abundance trends of the stars in Boxes D and F
should be similar to Box E. They are not. Instead Box D (inner disk) appears to contain
only old stars with thick disk abundance ratios, and only very few young stars
with thin disk abundance ratios. The opposite is true for Box F (outer disk) which
mainly contains young stars with thin disk abundance ratios and only a very few old
stars with thick disk abundance ratios. This means that the relative properties
of the thin and thick disks changes when going from galactocentric
distances ($R_G$) of 6\,kpc to 8\,kpc to 10\,kpc, in the sense that the thin disk
dominates at large $R_G$ while the thick disk dominates at small $R_G$.
If this is a real signature or just an effect of the various biases we have in
our sample is difficult to say. We have been selecting stars on eccentric orbits in
order to probe the thick disk, but no distinction has been made on if their orbits
place them in the inner or outer disk. However, recent observations of distant K giants
indicate that the abundance trends of the outer disk is dominated by
stars with thin disk abundance ratios , and that this might be a consequence
that the scale length of the thick disk is significantly shorter than that
of the thin disk \cite{bensby2010letter,bensby2011letter}. 
What we see in Boxes D and F could be a manifestation  of that.
If the thick disk scale length is shorter than the thin disk, we expect the thin
disk to dominate more and more with galactocentric distance.

\subsection{Formation of the thick disk}

A number of formation scenarios for thick disks have been proposed, e.g.: 
heating by mergers \cite{quinn1993,villalobos2010};
accretion of satellites \cite{abadi2003}; 
early, clumpy rapid star formation \cite{bournaud2009}; 
and migration \cite{loebman2011}. Although, it is likely that all of these
processes to some extent act in the Milky Way, it is not clear 
which if any is the dominant mechanism.

Thick disk formation scenarios, together with the observational 
constraints imposed from our spectroscopic survey, will be discussed in detail 
in an upcoming paper (Feltzing, Bensby, \& Oey, in prep.).
A full description of the sample of 703 stars, including abundances for
O, Na, Mg, Al, Si, Ca, Ti, Fe, Ni, Zn, Y and Ba, will be presented in another 
upcoming paper (Bensby, Feltzing, \& Oey, in prep.).

%
\bibliographystyle{epj}
\bibliography{referenser}

\end{document}